\documentclass[prl,preprint]{revtex4}

\usepackage{graphicx}
\usepackage{amssymb, amsmath}
\usepackage{bm}

\begin{document}

\title{Nonreciprocal propagation of surface acoustic wave in  Ni/LiNbO$_3$}
\author{R. Sasaki, Y. Nii, Y. Iguchi, and Y. Onose}
\affiliation{Department of Basic Science, University of Tokyo, Tokyo 153-8902, Japan}

\begin{abstract}
	We have investigated surface acoustic wave propagation in Ni/LiNbO$_3$ hybrid devices.
	We have found the absorption and phase velocity are dependent on  the sign of  wave vector  in a device, which indicates the nonreciprocal propagation characteristic of systems with time reversal and spatial inversion simultaneously broken  symmetries.
  	The nonreciprocity is  reversed by the 180$^\circ$  rotation of  magnetic field.
	Nonreciprocity seems largely dependent on the shape of ferromagnetic Ni film.
	The origin of these observations is ascribed to film shape dependent  magnetoelastic coupling.
	
\end{abstract}
\maketitle
	Breaking of spatial inversion symmetry (SIS) and time-reversal symmetry (TRS) largely influences on the spin dependent energy band dispersion of electrons $\epsilon({\bm k}, \sigma)$, where $\sigma=\uparrow,\downarrow$ is the spin state of electron. 
	When both the symmetries are preserved, $\epsilon ({\bm k}, \uparrow)=\epsilon ({\bm k}, \downarrow)= \epsilon (-{\bm k}, \uparrow)=\epsilon (-{\bm k}, \downarrow).$ If only TRS is broken, $\epsilon ({\bm k}, \uparrow)= \epsilon (-{\bm k}, \uparrow)$ but $\epsilon ({\bm k}, \uparrow) \neq \epsilon ({\bm k}, \downarrow)$. 
	The energy difference corresponds to the spin splitting in ferromagnetic metals. 
	For the systems with TRS but without SIS, $\epsilon ({\bm k}, \uparrow)=\epsilon (-{\bm k}, \downarrow)$ but $\epsilon ({\bm k}, \uparrow) \neq \epsilon (-{\bm k}, \uparrow)$ when ${\bm k}$ is along some specific crystal axis. 
	Rashba \cite{Rashba} effect and Dresselhause \cite{Dresselhaus} effect are examples of this type of spin splitting. 
	The degeneracy of $+{\bm k}$ and $-{\bm k}$ electronic states is completely lifted by the breaking of both the symmetries.

	The symmetry dependence of energy dispersion should be common to other elementary excitations in solids such as photon, magnon, phonon, etc. 
	For the photonic case, the circular polarization corresponds to spin state and slope of energy dispersion is inverse of refractive index. 
	In this sense, the natural optical activity and Faraday rotation are corresponding to the spin splitting in SIS-broken and TRS-broken electronic systems, respectively. 
	Recently, nonreciprocal directional dichroism in the systems without TRS and SIS has been investigated extensively \cite{Rikken,Kubota}. 
	This phenomenon is caused by the difference between the reflective indexes for +{\it k} and -{\it k} photons irrespective of the polarization direction, which originates from the photonic energy dispersion asymmetry in TRS-broken and SIS-broken systems. 
	Quite recently, similar nonreciprocal propagation of magnons has been observed in SIS-broken ferromagnetic films \cite{Zakeri, Zhang} and chiral ferromagnets \cite{Iguchi,Seki} originating from Dzyaloshinskii-Moriya interaction. 
	Here, we study the nonreciprocal propagation of phonon, more specifically, the surface acoustic wave in SIS- and TRS-broken hybrid devices.

	The surface acoustic wave (SAW) is the acoustic wave propagating on surface (Rayleigh wave). 
	The amplitude exponentially decays in the interior of material. 
	It has circular type polarization and the polarization direction is reversed upon the reversal of wave vector direction. 
	The SAW can be efficiently excited and detected by the electromagnetic wave on piezoelectric materials such as LiNbO$_3$ with use of interdigital transducers (IDTs) \cite{Datta}. 
	Recently, Weiler and coworkers have reported the  the ferromagnetic resonance can be excited by the surface acoustic wave on the ferromagnetic Ni/LiNbO$_3$ hybrid devices  \cite{Weiler,Dreher2012}.
	It should be noted that surface is obviously SIS-broken system and TRS is broken by the ferromagnet, and therefore the nonreciprocal propagation of SAW is expected in such a device. 
	We here study the SAW propagation on similar Ni/LiNbO$_3$ hybrid devices.  
	We show the SAW propagation is certainly nonreciprocal for one of the device. 
	Based on the device-dependence, we infer that the interference of  shear-type and longitudinal-type magnetoelastic couplings plays important role for the nonreciprocal propagation of surface acoustic wave.

	\begin{figure}
		\begin{center}
		\includegraphics[width=8cm]{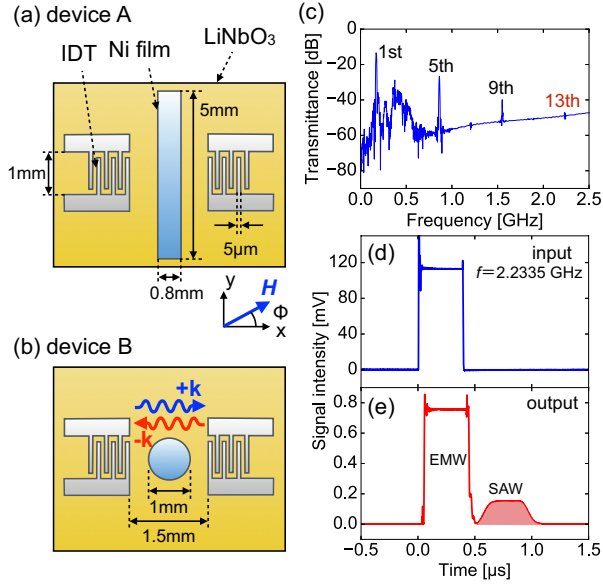}
		\caption{(a),(b) Schematics of two Ni/LiNbO$_3$ hybrid SAW devices used in this study. Rectangular ferromagnetic Ni film is evaporated on (a) deviceA and circular film  on (b) device B. (c) Typical transmission spectrum of device A. The 13 th harmonics of approximately 2.23 GHz is utilized in this study.
		(d),(e) Amplitude oscillograms for (d) input and (e) output signals at $f$ = 2.2335 GHz. Note that SAW signal and EMW noise are successfully separated in time domain using microwave homodyne circuit.}
		\end{center}
	\end{figure}
	To excite a high frequency SAW which couple to ferromagnetic resonance, we prepared SAW devices composed of two micro-fabricated IDTs and ferromagnetic Ni thin film evaporated on LiNbO$_3$ substrate as shown in Figs. 1(a) and (b). 
	We fabricated IDTs by photolithography technique and electron beam evaporation of aluminum. 
	The IDTs finger length and width  are 1 mm and 5 $\mu$m.
	The space between fingers is also 5 $\mu$m.
	The number of fingers for a IDT is 30. 
	Y-cut LiNbO$_3$ substrate was used and SAW propagation direction is along the Z direction of LiNbO$_3$.  
	The fundamental frequency of the SAW is estimated to be $\simeq$ 170 MHz \cite{Datta}.
	Ferromagnetic nickel film with the thickness of 30 nm is sputtered onto LiNbO$_3$ substrate between two IDTs. 
	We fabricated two devices with different shape of Ni films; one is rectangle (named device A)  and the other is circle (named device B) as shown in Figs. 1(a) and 1(b).
	
	We show a microwave transmission spectrum between two IDTs for the device A measured by a vector network analyzer ( Agilent E5071C ) in Fig 1(c).
	We observe a sharp peak around 170 MHz being consistent with the designed fundamental frequency of IDTs.
	In addition, 5 th,  9 th and 13 th harmonic peaks are also discerned. 
	In this study we use 13 th harmonic resonance ($f \simeq 2.24$ GHz), which is the highest in our device and comparable with the ferromagnetic resonance frequency of Ni film (see supplemental information \cite{Suppl}).
	 Precisely speaking, the excitation frequencies are 2.2335 GHz for device A and 2.2350 GHz for device B.
	There exists some direct electromagnetic crosstalk between IDTs  in addition to the coupling via SAW propagation.
	In order to separate the SAW signal from electromagnetic crosstalk, we perform a time domain measurement using pulsed microwave. 
	Figures 1(d) and 1(e) display demodulated microwave input and output signals through the SAW device.
	Using a large difference in group velocity between SAW and electromagnetic wave (EMW), we successfully separate them in time-domain.
	Our measurement system is based on a microwave homodyne circuit \cite{Luthi}, which allows simultaneous measurement of both amplitude and phase for transmitted SAW signal with high accuracy.
	All the experiments are carried out at 280 K.
	 In this experimental setup, the external magnetic field can be applied to any direction in SAW device plane, and the in-plane field direction is specified by the angle $\phi$  as shown in Fig. 1(a).
	
	\begin{figure}
		\begin{center}
		\includegraphics[width=8cm]{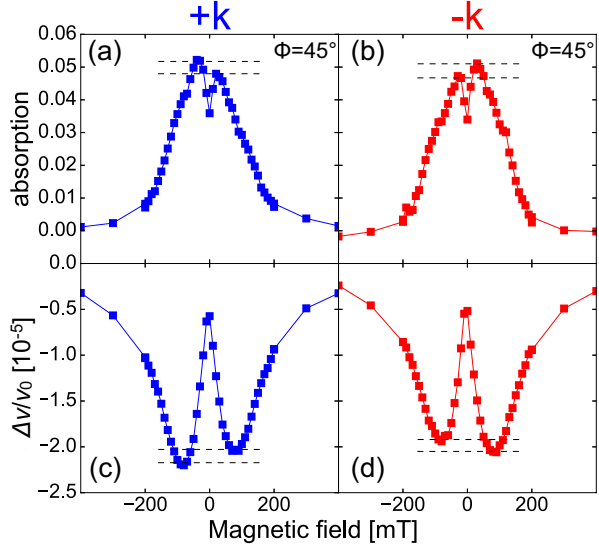}
		\caption{(a),(b) Absorption of SAW propagating along (a)  \lq\lq+{\it k}\rq\rq and (b) \lq\lq-{\it k}\rq\rq directions shown in Fig. 1(b) as a function of external magnetic field along $\phi = 45^\circ$ for device A. 
		(c),(d) Relative change of phase velocity of SAW propagating along (c)  \lq\lq+{\it k}\rq\rq and (d) \lq\lq-{\it k}\rq\rq directions  as a function of magnetic field along $\phi = 45^\circ$ for device A. 
		The dashed lines emphasize the asymmetry of magnetic field dependence in absorption and relative change of phase velocity.}
		\end{center}
	\end{figure}
	Figures 2(a) and 2(c) show the SAW absorption and the relative decrease of phase velocity for device A as a function of external magnetic field along $\phi = 45^\circ$.  
	Here, the absorption is defined by the decrease of transmittance normalized by the transmittance at high enough field (The absorption due to magnetoelastic coupling is minimal at high field as discussed below).
	The phase velocity is also normalized by the high field value. 
	The presented data are average of those taken by field increasing measurement and field decreasing measurement. 
	With decreasing magnetic field from +400 mT, the absorption increases largely.
	It shows a maximum accompanying a small dip structure around zero magnetic field.
	Then, the absorption decreases again toward high negative magnetic field. 
	Corresponding to the enhancement of the absorption, the phase velocity decreases in the low magnetic field region. 	Around the zero magnetic field, it shows a sharp peak corresponding to the dip structure of the absorption.
	The magnetization reversal process is restricted to the low field region below 30 mT as shown in supplemental material \cite{Suppl}.
	The dip structure of absorption and sharp peak of phase velocity are caused by the magnetization reversal process.
	On the other hand, the magnetic field dependences in the higher magnetic field region are induced by the change of difference in frequency between SAW and ferromagnetic resonance (FMR).
	While the SAW frequency does not  show magnetic field dependence, the FMR frequency increases linearly with the magnetic field.
	We certainly observe FMR around 2 GHz at zero magnetic field and its frequency increases with magnetic field for a similar Ni film (see supplemental material \cite{Suppl}).
	The energy difference tunes the magnetoelastic coupling, which is reflected by the absorption of SAW.
	
	Importantly we observe the difference of absorption between negative and positive magnetic fields, as emphasized by the dashed lines in Fig. 2. 
	Correspondingly, the phase velocity also shows the asymmetric field dependence.
	This asymmetries cannot be ascribed to the magnetization process because the magnetic moment saturates above 30 mT.
	Therefore the asymmetries should be caused by the intrinsic difference of SAW propagation between positive and negative magnetic fields, denoted as nonreciprocal SAW propagation.
	The SAW absorption $A(k,H)$  for wave vector $k$ at magnetic field $H$ satisfies the relation $A(k,H)=A(-k,-H)$.
	To check this relation, we observe the increase of absorption and the decrease of phase velocity for SAW with opposite wave vector (Figs. 2(b) and 2(d)). 
	As clearly shown in this figure the field asymmetries of absorption and phase velocity become reversed when the SAW propagates along the opposite direction.
	These results demonstrate the nonreciprocal SAW propagation in this Ni/LiNbO$_3$ hybrid device.

	\begin{figure}
		\begin{center}
		\includegraphics[width=8cm]{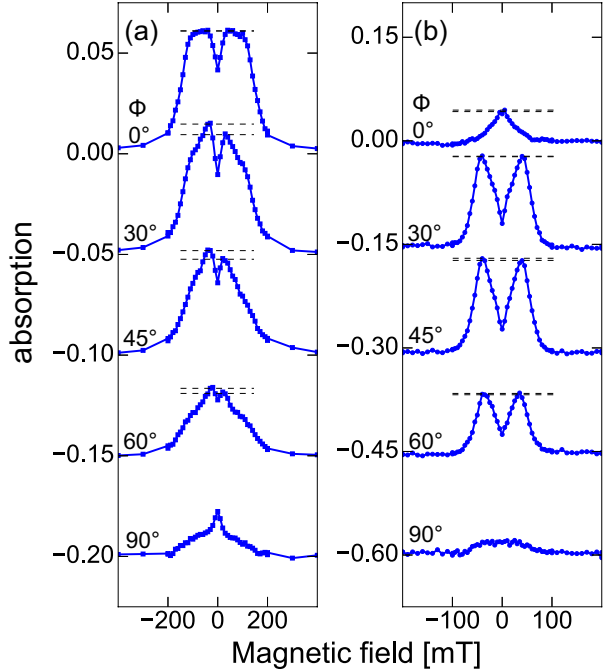}
		\caption{(a),(b) Magnetic field dependence of SAW absorption at several magnetic field directions for (a) device A and (b) device B. 
		The dashed lines emphasize the asymmetry of magnetic field dependence.}
		\end{center}
	\end{figure}
	\begin{figure}
		\begin{center}
		\includegraphics[width=8cm]{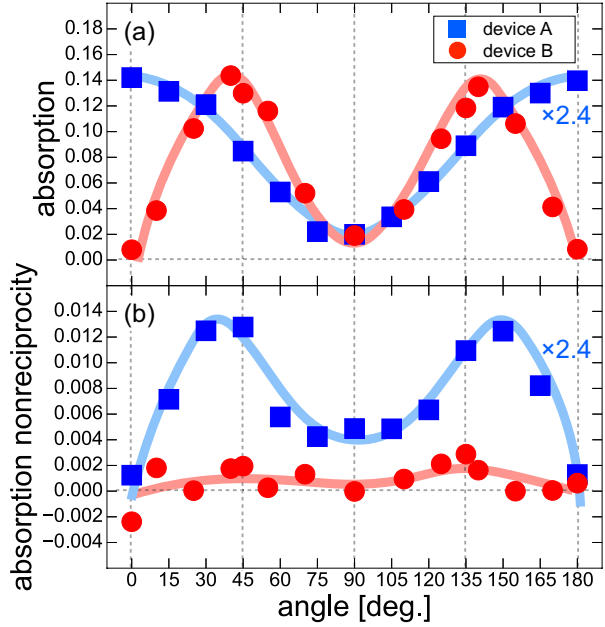}
		\caption{(a) The SAW absorption at $\mu_0H$ = 100 mT for device A and that at 40 mT for device B as a function of angle $\phi$. 
		(b)Averaged nonreciprocity as a function of angle $\phi$ for devices A and B. 
		The nonreciprocity is defined as the difference of absorption between positive and negative magnetic fields. 
		We averaged the device A data in the field region of 40 mT $\leq |\mu_0H| \leq$ 150 mT and device B data in the field region of 20 mT $\leq |\mu_0H| \leq$ 80 mT.
		Solid lines are merely the guide for the eyes.
		The absorption and nonreciprocity for device A are multiplied by 2.4 just for clarity.}
		\end{center}
	\end{figure}
	To study the nature of nonreciprocal SAW propagation, we investigate the magnetic field angle dependence of SAW absorption.
	We present the  SAW absorption at various $\phi$ for device A in Fig. 3(a).
	The absorption is large at $\phi = 0^\circ$ and decreases with $\phi$.
	We plot the absorption at  $\mu_0H$ = 100 mT as a function of $\phi$ for device A in Fig. 4(a).
	The absorption seems almost proportional to $\cos^2 \phi$.
	The nonreciprocity also shows a large angle dependence. 
	While the nonreciprocity is negligible at $\phi = 0^\circ$, the absorption at $\phi = 30^\circ$ clearly  shows the nonreciprocal behavior.
	The nonrciprocity decreases with increasing $\phi$ from $30^\circ$.
	The averaged nonreciprocity as a function of $\phi$ for device A is plotted in Fig. 4(b).
	The averaged nonreciprocity is defined as the averaged difference of normalized absorption between positive and negative magnetic fields in the  field region of 40 mT $\leq |\mu_0H| \leq$ 150 mT.
	The nonreciprocity steeply  increases with $\phi$ in low $\phi$ region.
	After showing the maximum around $\phi = 30^\circ$, it gradually decreases in the larger $\phi$ region.
	Nevertheless it does not vanish even at $\phi = 90^\circ$.
	Surprisingly the magnetic field dependences for device B are totally different.
	We show the SAW absorption at various $\phi$ for device B in Fig. 3(b).
	While the absorption at $\phi = 0^\circ$ and 90$^\circ $ is quite small, the large absorption is observed at 30$^\circ$, 45$^\circ$ and 60$^\circ$.
	The absorption is  suppressed above  100 mT.
	The suppression field is smaller than that of device A.
	The difference may originate from the difference of shape anisotropy.
	We also plot the angle dependence of the SAW absorption at 40 mT in Fig. 4(a).
	The absorption seems rather proportional to $\sin^2 2\phi$.
	More importantly the nonreciprocity is almost negligible in the case of device B.
	The magnetic field dependence of absorption is almost field symmetric as shown in Fig. 3(b).
	The averaged nonreciprocity in the field region of 20 mT $\leq |\mu_0H| \leq$ 80 mT plotted in Fig. 4(b) is also negligible.
	
	To discuss a microscopic origin of these experimental observations, let us theoretically discuss the SAW excitation of FMR.
	The lowest order of magnetoelastic coupling energy is \cite{Gurevich,Dreher2012}
	\begin{equation}
		F_{\rm coupling} = b_1 (m_x^2 \epsilon_{xx} + m_y^2 \epsilon_{yy} + m_z^2 \epsilon_{zz}) +2b_2(m_x m_y 		\epsilon_{xy} + m_y m_z \epsilon_{yz} + m_z m_x\epsilon_{zx})
	\end{equation}

	Here, $b_1$ and $b_2$ are magnetoelastic coupling constants for cubic symmetry and  $\epsilon_{ij}$ is $ij$ component of strain tensor.
	$\epsilon_{xx}$, $\epsilon_{zz}$ and $\epsilon_{zx}$ oscillate upon the SAW propagating along {\it x}-axis as followings \cite{Viktorov}; 
	\begin{eqnarray}
	\epsilon_{xx} = \epsilon_{xx0}\exp(i(kx - \omega t)),\\
	 \epsilon_{zz} = \epsilon_{zz0}\exp(i(kx - \omega t)),\\
	 \epsilon_{zx} = i\epsilon_{zx0}\exp(i(kx - \omega t)).
	\end{eqnarray}
	$\omega$ is SAW frequency, and $\epsilon_{xx0},\epsilon_{zz0}$, and $\epsilon_{zx0}$ are real and  decrease exponentially along {\it z} direction.
	Here, we assume that the device plane is perpendicular to {\it z}-axis.
	The phases of longitudinal strains $\epsilon_{xx}$, $\epsilon_{zz}$ and shear strain $\epsilon_{zx}$ are different by $\pi$/2 forming the ellipsoidal polarization of elastic oscillation.
	When $k$ is reversed, the ellipsoidal direction becomes reversed  and  the $\epsilon_{zx0}$  shows different sign.
	We also assume the static magnetization is parallel to  the device plane as followings;
	\begin{equation}
		\begin{pmatrix}
			m_x\\
			m_y \\
			m_z
		\end{pmatrix}
		=
		\begin{pmatrix}
			m_0\cos \phi \\
			m_0\sin \phi \\
			0
		\end{pmatrix}.
	\end{equation} 
	Then the effective magnetic field induced by SAW propagating along {\it x}-axis is
	\begin{equation}	\label{dynamical effective field}
		-\bm{\nabla_m}F_{\rm coupling} 
		=\mu_0 
		\begin{pmatrix}
			h_x\\
			h_y \\
			h_z
		\end{pmatrix}
		=
		\begin{pmatrix}
			2b_1m_x\epsilon_{xx} \\
			0 \\
			2b_2m_x\epsilon_{zx}
		\end{pmatrix}
		=
		\begin{pmatrix}
			2b_1 m_0 \epsilon_{xx0} \cos \phi \exp(i(kx - \omega t))\\
			0 \\
			2 i b_2 m_0 \epsilon_{zx0} \cos \phi \exp(i(kx - \omega t))
		\end{pmatrix}.
	\end{equation}
	Only the AC magnetic field perpendicular to the magnetization $h_\perp$ can excite FMR.
	The in-plane and out-of-plane  components of $h_\perp$ are $h_x\sin \phi$ and $h_z$, respectively.
	In addition,  the susceptibility of FMR for right-handed circular polarized field $\chi_+$ is large while that for left-handed field $\chi_-$ is negligible \cite{Gurevich} .
	The SAW absorption  induced by magnetoelastic FMR is proportional to 
	\begin{equation}   \label{absorption}
			\chi''_+ |h^+_\perp|^2 = \chi''_+ |h_x\sin \phi + i h_z|^2   
			= \chi''_+|2b_1m_0\epsilon_{xx0}\cos \phi \sin \phi + 2b_2 m_0 \epsilon_{zx0} \cos \phi |^2 ,
	\end{equation}
where $\chi''_+$ is the imaginary part of magnetic susceptibility for right-handed circular polarized field.
	As mentioned above, the SAW absorption for device A is roughly  proportional to $\cos^2 \phi$, while that for device B to $\cos ^2 \phi \sin^2 \phi$.
	These indicate that $b_2$ ($b_1$) contribution to the magnetoelastic excitation is dominant for device A (device B).
	The difference may originate from the shape dependent magnetoelastic coupling but at present we could not specify  the microscopic mechanism.
	The difference of nonreciprocity between devices A and B is also ascribed to the different magnitude of $b_1$ and $b_2$.
	When the wave vector of SAW is reversed, the circular direction of elastic wave polarization is reversed, which is reflected by  the sign of  $\epsilon_{zx0}$. 
	Therefore the nonreciprocity can be induced by the interference of $b_1$ and $b_2$ terms in Eq. (\ref{absorption}).
	In device B the minor $b_2$ term seems negligible and therefore the nonreciprocity cannot be observed.
	On the other hand, in device A,  the  minor $b_1$ term is still active so that the nonreciprocal SAW propagation is finite.
	
	In conclusion, we have successfully observed  nonreciprocal propagation of surface acoustic wave in a Ni/LiNbO$_3$ hybrid device. 
	Based on the device dependence of nonreciprocity, we infer that the film-shape dependent magnetoelastic coupling is crucial for the nonreciprocity.
	While nonreciprocal microwave devices have been fabricated from the ferromagnetic components, the surface acoustic wave devices have been  used mainly for  bandpass filters. 
	Our observation may pave a new route to nonreciprocal microwave device based on surface acoustic wave.
	
	The authers thank K. Ueno for technical assistance. This work was in part supported by the Grant-in-Aid for Scientific Research (Grants No. 25247058, No. 16H04008, and No. 15K21622) from the Japan Society for the Promotion of Science, Yamada Science Foundation, and The Murata Science Foundation.


\clearpage
\section*{Supplementary Information}
\subsection*{Magnetization curve}
	\begin{figure}
		\begin{center}
		\includegraphics[width=8cm]{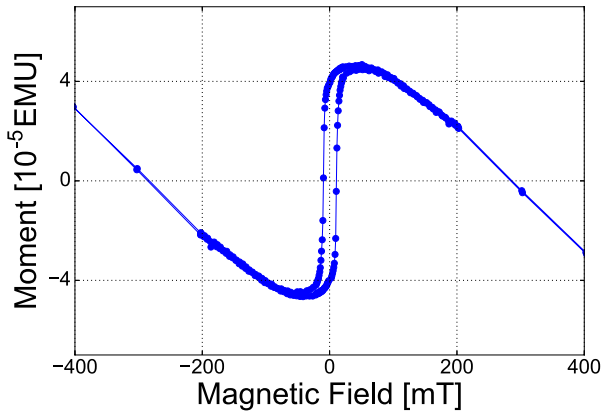}
		\caption{Magnetization curve of Ni film at 280 K}
		\end{center}
	\end{figure}
	We plot in Fig. 5 the magnetization curve of Ni film with the thickness of 40 nm measured by magnetic property measurement system (Quantum Design). The Ni film is sputtered on LiNbO$_3$ substrate. The magnetization is saturated at 30 mT. The moment is decreased at high magnetic field because of diamagnetic signal from the substrate.
\subsection*{Ferromagnetic resonance}
	\begin{figure}
		\begin{center}
		\includegraphics[width=8cm]{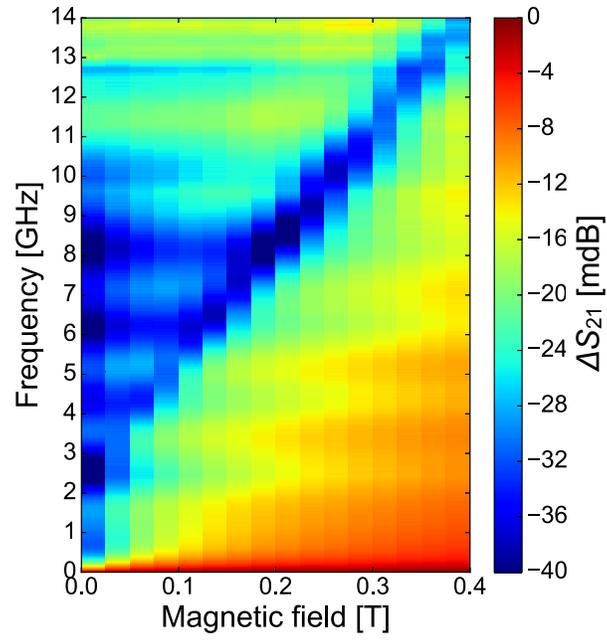}
		\caption{The contour plot of microwave absorption of Ni film in the frequency-magnetic field plane.}
		\end{center}
	\end{figure}
	In Fig. 6, we show the microwave absorption of Ni film at 280 K as  functions of frequency and magnetic field, which is measured in the measurement system  shown in a literature (ref. 7 of main text).
	The linear magnetic field dependence of the absorption peak frequency is clearly observed above 0.2 T. 
	The origin of absorption peak can be ascribed to the ferromagnetic resonance.
	If the linear magnetic field dependence is extended to zero magnetic field, the zero field frequency is around 2 GHz, which is comparable with the SAW frequency used in our measurement.
\end{document}